\documentstyle[12pt]{article}
\begin{document}
\baselineskip=14truept plus 0.2truept minus 0.2truept

\begin{center}

{\bf Do neutrinos have mass only within matter?}

\end{center}

\medskip

\begin{center}

R. F. Sawyer

\end{center}

\begin{center}

Department of Physics, 
University of California at Santa Barbara
\\
Santa Barbara, California 93106

\end{center}

\medskip

\begin{abstract}
 We look at the possibility that appreciable neutrino masses and flavor mixing occur only within material media, driven by an interaction between leptons and a very light scalar particle. Limits are placed on the scalar particle mass and coupling constants from a number of experimental and astrophysical considerations.
 \end{abstract}

\bf 1.Introduction. \rm
The recent results on atmospheric neutrinos \cite{fukuda}-\cite{becker} offer a persuasive argument that  neutrino mass and flavor oscillations are needed to resolve the anomalies with respect to standard model
predictions. Once one accepts this conclusion, a parallel explanation of the solar neutrino deficits seems inevitable. Nevertheless, the evidence that neutrinos in a vacuum
have mass or oscillate is still inconclusive. In particular, in the vacuum oscillation model that has been use to fit the data in \cite{fukuda}, except for the predictions for the lowest energy bin, the flavor mixture of neutrinos from above is changed little from that at the neutrino production point, while the flavor mixture of upcoming $\mu_\nu$ is changed greatly, by virtue of the much greater path lengths from the point of production. In the lowest energy bin, $p<.4 $ GeV/c ,  the directional determination appears to be too imprecise to test the premise that only the upmoving neutrinos are changed. It is therefore worth considering the possibility that the effects depend not just on the distance but on the intervening matter as well. Indeed, Wolfenstein  \cite{wolf} long ago discussed such a possibility, pointing out that experiments do not rule out flavor changing neutral currents, at essentially standard strength, that could cause neutrino flavor to precess in matter while not producing, or demanding, neutrino mass. 
 
In somewhat the same spirit we here suggest that the intervening matter induces effects that are tantamount to masses and flavor oscillations of the upmoving atmospheric neutrinos, and in the neutrinos from the sun as well. Since two particle cross-sections involving a neutrino and any constituent of the matter are far too small to make a
substantial effect on the neutrino fluxes, any such effect must come from beyond the standard model and also involve some kind of coherence, or other long range phenomenon. We consider here the case in which a light scalar particle
coupled very weakly to both neutrinos and to a constituent of the matter gives rise to neutrino masses and mixing within the medium. It is the small mass of the scalar particle that allows the very weak
coupling to produce the requisite mass differences, while not changing two particle cross-sections appreciably.  

We take a real scalar field, $S(x)$,with mass, $\mu$, interacting with electrons and neutrinos through the interaction Lagrangian density,
\begin{equation}
L_{I}= G_e\bar \psi_e \psi_e S + \sum_{i,j}G_{i,j}: \bar \nu_j \nu_i S: ,
\label{1}
\end{equation} 
where the indices, $i,j$ , run over the three flavors of massless Dirac neutrinos. We take the coupling matrix $G$ to be of rank one or two, so that there will always remain at least one neutrino combination that is uncoupled to S. We have not prescribed the couplings of the scalar field S to $\mu^{\pm}$, $\tau^{\pm}$, since they will not enter into the considerations that follow. For the case of  an otherwise free S field, the interaction (\ref{1}) leads in  lowest order to a mass matrix, $m_{i,j}$, for neutrinos in a medium of nonrelativistic electrons with electron density $n_e$,  
\begin{equation}
m_{i,j}=G_e G_{i,j}n_e\mu^{-2} .
\label{2}
\end{equation}

Since the symmetry, $ \nu \rightarrow \gamma_5\nu$ , $S\rightarrow -S$,  is broken by the coupling to the electrons, loops involving electrons will generate a vacuum neutrino mass, beginning in third order in the parameter $G_e $.  However we shall take $ G_e$ to be extremely  small, and the vacuum masses should be negligible.
We can choose the couplings and S particle mass to make a variety of oscillation-mixing models.  We should mention that similar ideas have been presented in refs. \cite{steph} and \cite{horvat}. In \cite{steph} the scalar field is taken to couple only to $\nu$'s, which have a vacuum mass of some other origin, with the possibility of clustering to produce effects somewhat like ours. In \cite{horvat}, in which the scalar particle is a dilaton, the $\nu$ mass is given by an expression analogous to (\ref{2}). Before giving examples in our model we consider constraints on the parameters that must be imposed to avoid contradictions with what is already known.

\bf 2. S particle luminosity of the sun. \rm Since the S particles will have mass of much less than one KeV, they can be produced thermally in the sun; and since their interactions with ordinary matter will be far too weak to generate S-opacity in the medium, the interaction must be weak enough that the S luminosity of the sun, $L^{(S)}_{sun}$ is a small fraction of the solar constant. We estimate this luminosity by considering the free-free, free-bound, and bound-bound processes of S emission and absorption, as in the photon case. For both photons and S particles we use the lowest term in the expansion in powers of k of the angle averaged (and polarization summed, for the photon) squared matrix element between nonrelativistic electronic states, where $(\omega,k)$ is the energy-momentum of the absorbed particle. For a photon it is \hspace{.1 in}$(2/3)e^2\omega^2|<f|\vec r|i>|^2$; for the scalar particle it is $(1/3)G_e^2 k^2|<f|\vec r|i>|^2$. Thus for energies of a few KeV, where $k\approx \omega $ for the S, the inverse absorption lengths $l_\gamma^{-1}$, and $l_S^{-1}$ are related by the coupling constant ratio.
Detailed balance  turns $l_\gamma^{-1}$ into a rate of photon energy emission per unit volume. We then scale this rate down by the coupling constant ratio. Using median values
$l^{-1}_\gamma \approx .003$ cm., $T\approx .8$KeV, and radius of $5 \times 10^{10}$ cm. \footnote{We used the opacities plotted as a function of $T$ and $\rho$  in D. D. Clayton, \it Principles of Stellar Evolution and Nucleosynthesis \rm McGraw-Hill, New York (1968). A multizone estimate using data from these figures gives nearly the same limits on coupling. There is a similar scaling of the rate for $\gamma+e^- \rightarrow S + e^-$ to the Compton scattering rate in the star, which enables us adequately to take into account this process. }, we estimate a rate of, $L^{(S)}_{sun}\approx 2\times 10^{58}G_e^2$ ergs/sec, to be compared with the solar constant, $3.90\times 10^{33}$ ergs/sec, leading to the bound, $G_e^2<<10^{-25}$. This is a smaller bound on $G_e^2$ than that found for the axion coupling in the estimation of light axion bremsstrahlung luminosities \cite{krauss}, by a factor of about $10^6$, for the reason that pseudoscalar coupling in the axion case produces an extra factor of  roughly $T^2/m_e^2$ in the rate.  Our combined  processes have temperature dependence  roughly proportional to $T^{1/2}$; the weak dependence on temperature makes it unlikely that hotter stars will impose more severe restrictions, in contrast to the axion case.

\bf 3. Long range forces \rm  For a massless S and a value of $G_e^2=10^{-25}$ the long range attraction 
between lumps of ordinary matter  would be roughly $10^{12}$ times that of gravity. For $\mu>10^{-2}$eV, or the range of the force less than .002 cm., there appears to be no conflict with long range force measurements. However it will turn out that these constraints put us on the very border of the domain of values that are needed for atmospheric neutrino transformations. \footnote{The constraints could be relaxed through the agency of an S self-interaction potential, which can change the connection between the S field inside a medium and the long range force between two separated lumps of matter.} Defining the model above, with $G_e^2<10^{-25}$ and $\mu>10^{-2}$ eV, as ``weak", we also consider a less constraining ``very weak" form, in which
$G_e^2<10^{-43}$ and $\mu>10^{-11}$eV; or a range up to 20 km and a strength less than $10^{-6}$ that of gravity, putting the parameters well within the bounds allowed by the ``fifth force" searches in recent years \cite{adelberger}. 

\bf 4. Atomic physics. \rm  The limit on the electron-S coupling in sec.2 guarantees that the forces from S exchange make corrections much less than one part in $10^{-23}$ to the energy levels of the positronium system and negligible contributions to the binding energy per electron of bulk matter.

\bf 5. Supernova neutrino pulse--transparency of the interstellar medium. \rm The most stringent bounds on $G_{\nu}$ are set by the
observations of neutrinos from SN 1987a.  It has been shown \cite{kolb} that the absence of large
scattering from relic big bang neutrinos (of SN neutrinos coming from the LMC) can be assured with a coupling $G_{\nu}^2<10^{-7}$. Thus the upper limit of the combination that enters our $\nu$ masses is $G_{\nu}G_e \mu^{-2}<10^{-10}$(eV)$^{-2}$ for the ``weak" choice,  and  $G_{\nu}G_e \mu^{-2}<10^{-3}$(eV)$^{-2}$ for the ``very weak" choice of parameters. Although we choose not to, we could avoid even this limitation, since we are taking at least one combination of neutrinos not to couple to S. As long as we have a single uncoupled $\bar\nu$ combination that contains a reasonably large fraction of $\bar\nu_e$ the couplings are not limited by the data from 1987a.

\bf 6. Neutrino cross-sections.  \rm   We demand that the electron-neutrino cross-sections produced by S exchange between $e_-$ and $\nu$ , be significantly smaller than standard model cross-sections, for energies down to 1 MeV. Taking $G_\nu$ to be the maximum eigenvalue of the coupling matrix, and taking only the partial cross-section for angles greater than some small minimum gives,
\begin{equation}
G_e^2G_\nu^2 < G_F^2(1 MeV)^4\approx 10^{-22},
\label{3}
\end{equation}
which is satisfied by many orders of magnitude in either of our parameter domains. The small angle part gives a correction which is of order $m_{\nu}/\mu$ times the above, which again cannot be appreciable.

\bf 7. Supernova neutrino pulse--generation. \rm If we begin with a coupling strength and range that gives a mass of $10^{-2}$eV for one of the neutrino states in ordinary matter, and if
in going to superdense matter at, say, $10^{12}$ g/ cm$^{3}$, the mass scales up according to ( \ref{2}), then the neutrino mass in the dense medium would be at the GeV level. This cannot be the correct complete description in a domain in which the coupling has become effectively so strong, but it does appear that in the superdense case one or two of the neutrino combinations would not be available to participate in the job of energy and lepton number transport to the surface. However, it was shown in \cite{saw} that in the denser parts of the supernova core quite excessive neutrino opacities have been used in the calculations of the pulse, due to the neglect of interaction effects in the medium. In consequence, the rate of core energy flow of the current calculations can be realized with fewer than three neutrino species, without changing the temperature profile by much. Add to this the ability of the system to adapt by changing the temperature profile, and we get a picture with a slightly larger and hotter neutrinosphere than in present models \footnote{It is larger because at the lower densities of the neutrinosphere the higher energies of the neutrinos emerging from the dense core count more than the interaction-induced reduction in cross-sections in determining the interaction length.}. The temperature would decrease more slowly than in the three neutrino model, since there would be only one or two neutrino combinations excitable, even at the neutrinosphere densities. As for the effect of $\nu-\nu$ scattering on the transport of energy out of a neutron star, we provisionally agree with the considerations in \cite{dicus} indicating that it will not make a drastic difference. In the region in which the neutrinos move out diffusively we can use a frame that is moving at the drift velocity to see that in this frame $\nu$-$\nu$ scattering does not change the distribution functions. However, for the larger ranges of our $\nu$-S coupling the $\nu$-$\nu$ scattering in the interior will create a quasi-equilibrium of otherwise sterile $\nu_R$'s, which are effectively required by the $\nu$-$\nu$ interaction to drift at the same speed as the $\nu_L$'s. In this case, the pulse would be only 50\% non-sterile, at best. If this is a problem, it can be completely removed by choosing the Majorana variant which we mention later. Another potential problem is an effect on the neutrino spectrum as we go from the diffusive region to the free-streaming region. We expect this to be small, since an element of neutrino gas does not expand very much in the rather narrow transition region.  

\bf  8. Coupling schemes. \rm  We treat the simulation of the mixing
scheme in which the atmospheric neutrino data are interpreted through pure $\nu_\mu-\nu_\tau$
oscillations \footnote{ Ref.\cite{barger} discusses the amount of mixing of  $\nu_\mu$ with $\nu_e$  allowed in three neutrino oscillation schemes that fit the SK atmospheric data and are compatible with the CHOOZ data \cite{chooz} with  the conclusion that an 18 \% admixture in amplitude is allowed. Although we do not explore such possibilities, our class of models should allow a considerably larger admixture, since the electron density along the path of CHOOZ neutrinos is roughly 1/2 that of the earth's mantle.}and the solar neutron problem is addressed through mixing of $\nu_e$ and some other flavor (or flavor combination). We take the eigenstates of the matrix
$G_{i,j}$ to be the following combinations
\begin{eqnarray}
\nu_1=\cos(\alpha)\nu_e+2^{-1/2}\sin(\alpha)[\nu_\mu+\nu_\tau],
\nonumber\\
\nu_2=-\sin(\alpha)\nu_e+2^{-1/2}\cos(\alpha)[\nu_\mu+\nu_\tau],
\nonumber\\
\nu_3=2^{-1/2}[\nu_\mu-\nu_\tau],
\label{4}
\end{eqnarray}
with the respective coupling eigenvalues $G_{1,2,3}$ and mass eigenvalues (in the medium)
$m_{1,2,3}=G_e G_{1,2,3}\mu^{-2}n_e$. We take $G_3>>G_2$ and  $G_1=0$. Beginning with the atmospheric case, we note that the only mixing induced by $G_3$ alone is the mixing between $\mu_\nu$ and $\mu_\tau$ with a maximal mixture, $\sin(2 \theta)=1$, and a mass difference given by $\delta m \approx m_3$, which we shall take in the region,
$10^{-1}$-$10^{-2} eV$. Using mantle electron densities in the earth of $10^{24}$ cm$^{-3}$ in (\ref{2}) this translates  into
\begin{equation}
G_3G_e(\mu/1 eV)^{-2} \approx [6\times10^{-12}-6\times10^{-13}],
\label{5}
\end{equation}
which is satisfiable, with rather little leeway for the ``weak" model, but with much to spare in the ``very weak" model.

The neutrino combination, $\nu_3$, does not enter the solar problem; only the symmetric combination, $\nu_\mu-\nu_\tau$ mixes with the electron neutrinos. 
For a particular $E_\nu$ we can choose $G_2$ and $\alpha$ to produce a resonance at some particular point in  the sun. But higher energy $\nu$'s will be resonant at higher densities and lower energy $\nu$'s at lower densities, because of the density dependence of the mass matrix. In a small angle MSW model, fits to data are achieved in which the survival probability is almost nil at $E_\nu= 1 MeV$ but increases rapidly as the energy is lowered to the detector threshold, because of the resonance region moving into the energy producing region in the sun. The survival increases, but more slowly, as the energy is increased from 1 MeV, this time because of the loss of adiabaticity. The effect of this loss can be estimated quite accurately using the Landau- Zener type calculation of the transition probability between the two eigenstates \cite{haxton}. In a small angle model in which $G_2$ has been set to agree with  $\delta m^2$ from MSW at some intermediate density, this transition probability for our model is the same function of the resonant density and mass-matrix mixing angle as in MSW. But in contrast to the MSW case, it becomes significant at low energies, while the restoration of survival at higher energies can be achieved by the resonance moving into the energy producing region. It appears that the existing data is less well fit by such a model, with its gentler rise to the left of the minimum and sharper rise to the right, than it is by small angle MSW. 

However, in our model a large angle example can provide the following: a survival fraction, $f_s$,  of about .4, at E=.4 MeV, which increases almost linearly to a value $f_s \approx .5-.6$ at E=15 MeV, depending on choice of parameters, in the absence of regeneration in the earth. This is achieved with $.62< \sin\alpha<.68$ and $4\times 10^{-14}<(\mu/1eV)^{-2}G_2G_e<8\times 10^{-14}$. For these parameters, nighttime regeneration can give an increase of $f_s$ in the region $E<$2MeV,  reaching a 25\%-30\%  increase in $f_s$ at .4 MeV. The all-over behavior is qualitatively consistent with the survival curves for the preferred MSW models given in  \cite{bahcall}, fig. 9.

 \bf 9. Early universe, just before nucleosynthesis. \rm  Here the medium consists almost entirely of electrons, positrons, neutrinos, antineutrinos, and photons. The interaction (\ref{1}) provides long
range interactions among all of these species except the photons. The induced mass, $m_3$, of the most strongly coupled $\nu$ combination, coming from interaction with the thermal electrons and positrons is, in lowest order,
\begin{equation}
m_{3}=4(2\pi)^{-3}G_3 G_e m_e \mu^{-2}\int \frac{d^3p}{\sqrt{p^2+m_e^2}}
[1+\exp(\sqrt{p^2+m_e^2}/T)]^{-1} ,
\label{6}
\end{equation}
Taking $G_3G_e \mu ^{-2}$ from (\ref{5})we find, for example, that in the electron-positron plasma at T=2 MeV the range of $m_3$ is .64-6.4 MeV, so that there would appear to be the possibility of largely excluding or largely allowing a sea of $\nu_3, \bar\nu_3$ during this critical period in the expansion. In the early universe there is not sufficient excess of electrons over positrons to mix the states (\ref{4}) by a significant amount and the mass $m_2$ induced when $G_3 \rightarrow G_2$ in (\ref{6}) will be much smaller. A perturbative estimate of the mass changes induced by $\nu-\nu$ interactions, through (\ref{6}) with $m_e\rightarrow m_\nu$, will give small changes, owing to the neutrino mass factor on the right hand side. \footnote{But nonperturbative solutions of the coupled  $\nu-\nu$ mass equations might lead to much larger masses than those discussed above for both $\nu_2$ and $\nu_3$.} We conclude that there is reduction in the number of effective left-handed $\nu$ species,  most probably from 3 to 2, but possibily greater or less.  

There are also new degrees of freedom excited in the soup. S particles equilibrate easily from the reaction $\nu+\bar\nu\rightarrow 2S$, and right handed $\nu$'s are produced easily in  $\nu^L+\bar\nu^L\rightarrow \nu^R+\bar\nu^R $, giving effectively, 4.5 species. This can be reduced  to a maximum of 2.5 by using Majorana fields for the neutrinos, with the scalar coupling the analogue of a Majorana mass term.

\bf 10. Conclusion. \rm Until such time as the data rule out our initial qualitative assumption, that the downgoing atmospheric $\nu$'s are not altered from the point of production, it is worth
considering the possibility that neutrinos remain nearly massless and unmixed in the vacuum. As to the class of models proposed here, the largest negative aspect is the introduction of the new particle. The possibility of reducing the effective number of neutrinos in the early universe may be a positive aspect, and the extension of the duration of the supernova pulse, without diminishing the effective temperature, may be one as well.

\end{document}